\documentclass[aps,prd,twocolumn,nofootinbib,superscriptaddress,10pt,floatfix]{revtex4-1}
\usepackage[T1]{fontenc}
\usepackage{amsmath,amssymb,graphicx,booktabs,microtype,hyperref,placeins}
\hypersetup{colorlinks=true,linkcolor=black,citecolor=black,urlcolor=blue}
\newcommand{\Md}{M_{\rm d}}
\newcommand{\Ma}{M_{\rm a}}
\newcommand{\Om}{\Omega}
\newcommand{\K}{\mathcal{K}}
\newcommand{\Dt}{\mathcal{D}}
\newcommand{\Aa}{\mathcal{A}}
\newcommand{\Mo}{M_\odot}
\begin{document}
\title{The Stability Frontier of Interacting Binary Stars}
\author{Hisham Anwer}
\affiliation{Physics Program, Zewail City of Science, Technology and Innovation, Giza, Egypt}
\affiliation{Department of Physics, Faculty of Science, Cairo University, Giza, Egypt}
\email{hisham@cantab.net}
\begin{abstract}
We derive a conservation based dynamical stability frontier for interacting binary stars that distinguishes stationary mass transfer from the restoration of a perturbed contact state. We apply the formalism to conservative white dwarf binaries with direct impact accretion, coupling Roche overflow, stream angular momentum, gravitational radiation, and both stellar spin reservoirs. From the contact equations, we obtain the required orbital return torque, and an exact Jacobian identity yields the restoring gradient and an oscillatory stability boundary in the computed white dwarf domain. A dimensionally normalized quadrupolar tide reveals separate equilibrium fold and oscillatory thresholds, including coexisting stable and unstable stationary branches. Finite donor synchronization introduces an additional oscillatory stability frontier. The model recovers the classical Roche and gravitational radiation limits, while the calculated orbital period derivatives are consistent with published binary timing measurements within their combined observational and model uncertainties. These results specify the torque amplitude, restoring stiffness, and synchronization rate needed for local stability, providing quantitative targets for stellar tidal calculations and the interpretation of interacting binaries in the millihertz gravitational wave band.
\end{abstract}
\maketitle
\section{Introduction}
Mass transfer transforms the evolution of interacting binary stars by redistributing angular momentum between the orbit, the accretion flow, and the stellar spins. The donor's structural response and the changing Roche surface determine the onset of contact, while the response of the coupled system to perturbations determines its dynamical survival. A stability theory must therefore resolve both stationary mass exchange and the restoring dynamics that sustain it.

The evolution of binary stability theory reflects a fundamental progression from locating Roche contact to understanding the dynamical processes that sustain it. Stellar mass and radius relations, together with Roche geometry, establish the structural response of a mass losing donor and the evolution of its critical surface~\cite{Nauenberg1972,Eggleton1983}. Classical overflow prescriptions and subsequent hydrodynamic nozzle calculations connect this geometry to the mass transfer rate and its sensitivity to contact depth~\cite{Ritter1988,Cehula2023}. The role of angular momentum transport became central when Marsh, Nelemans, and Steeghs established the influence of direct impact accretion and accretor synchronization on mass transfer stability~\cite{Marsh2004}. Gokhale, Peng, and Frank extended the dynamical description to donor spin and oscillatory mass transfer~\cite{Gokhale2007}. Subsequent investigations of stellar tides revealed resonant, traveling wave, and nonlinear torque responses whose strength depends sensitively on the forcing frequency~\cite{Burkart2013,FullerLai2014,Yu2020}. These advances establish the physical ingredients for a unified stability criterion in which the torque sustaining stationary contact and the restoring response to perturbations are evaluated at the same binary configuration.

Building on these developments, we introduce a conservation based dynamical model that derives the orbital return torque required at stationary contact and the independent torque gradient governing local restoration. The coupled equations incorporate the nozzle response, evolving stellar moments of inertia, and both stellar spin reservoirs. Our formalism explicitly recovers the classical conservative Roche, direct impact, and gravitational radiation limits~\cite{Eggleton1983,Marsh2004,Gokhale2007}. We derive an exact Jacobian determinant identity and the associated oscillatory stability criterion, then evaluate a dimensionally normalized quadrupolar tidal torque to establish distinct thresholds for stationary branch existence and dynamical stability, connecting these requirements to stellar tidal response calculations~\cite{Burkart2013,FullerLai2014,Yu2020}. By extending the dynamical equations to finite donor synchronization, we derive an additional oscillatory frontier. We calculate direct impact white dwarf benchmarks and orbital period derivatives, obtaining consistency with independent binary timing measurements~\cite{Ramsay2018,Green2018,Green2026}. For the reference $0.20+0.60\,\Mo$ binary, the orbital period $P=216.6\,\mathrm{s}$ corresponds to $f_{\rm GW}=9.23\,\mathrm{mHz}$, connecting the calculated stability frontier to the population of interacting binaries accessible to millihertz gravitational wave observations~\cite{Toubiana2024}. The foundation of these results is a unified description of Roche contact and angular momentum exchange, which we now construct from the orbital dynamics and the evolving spins of both stars.

\section{Roche Contact and Spin Angular Momentum}
The restoration of mass transfer depends on the evolution of the donor's Roche surface and the redistribution of the stream's angular momentum. We begin by combining Roche geometry with a conserved orbital and spin angular momentum budget. For donor and accretor masses $\Md,\Ma$, define $q=\Md/\Ma$, $M=\Md+\Ma$, the separation $a$, the reduced mass $\mu_{\rm red}=\Md\Ma/M$, the orbital momentum $J=\mu_{\rm red}\sqrt{GMa}$, and the orbital frequency $\Om=(GM/a^3)^{1/2}$. Newton's constant is $G$. Writing $R_L=af(q)$ and $R_{\rm d}$ for the donor radius, let $f_q\equiv d\ln f/d\ln q$ and $\zeta_{\rm d}\equiv d\ln R_{\rm d}/d\ln\Md$. Eggleton's geometry gives~\cite{Eggleton1983}
\begin{equation}
\begin{aligned}
 f(q)&=\frac{0.49q^{2/3}}{0.6q^{2/3}+\ln(1+q^{1/3})},\\
 \K&=\zeta_{\rm d}+2(1-q)-(1+q)f_q.
\end{aligned}
\label{eq:roche}
\end{equation}
\begin{figure}[t]
\centering
\makebox[\columnwidth][c]{%
  \vbox{%
    \hbox{\includegraphics[width=.96\columnwidth]{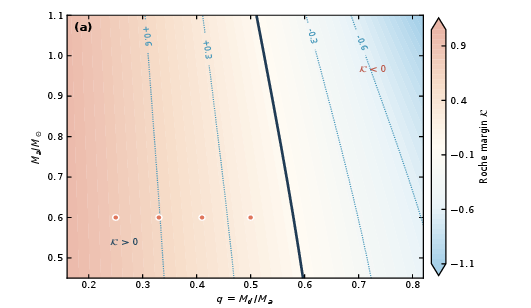}}%
    \nointerlineskip
    \hbox{\includegraphics[width=.96\columnwidth]{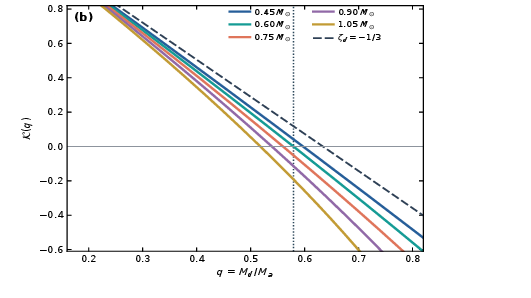}}%
  }%
}
\caption{Structural Roche benchmark recovered as the zero-spin, fully returned-stream limiting case. Panel (a) retains the multi-contour Eggleton--Nauenberg stability map; panel (b) shows five accretor-mass sequences and the constant-slope donor benchmark. The exact crossing at $q_{\rm crit}=0.5793833$ for $M_{\rm a}=0.60M_\odot$ and the fixed-slope limit $0.6339795$ follow directly from the conservative Roche geometry. The subsequent contact calculations include the donor spin correction.}
\label{fig:roche_v20}
\end{figure}

The structural boundary $\K=0$ reproduces the classical conservative Roche limit when the stream returns its angular momentum and stellar spin contributions vanish. We define $m=-\dot\Md=\dot\Ma>0$ denote conservative mass transfer, $j$ the net stream specific angular momentum delivered to the accretor, and $N(\delta)>0$ the return torque on the orbit, where $\delta=\omega_{\rm a}-\Om$ is the accretor's spin lag. Gravitational radiation removes angular momentum at rate $\gamma_{\rm GW}J$, with $\gamma_{\rm GW}>0$.

The accretor and synchronized donor carry spin moments $I_i=k_i^2M_iR_i^2$ for $i={\rm a,d}$, with $\omega_{\rm d}=\Om$. Write $I'_i=dI_i/dM_i$, $\eta_i=d\ln I_i/d\ln M_i$, and $\epsilon_i=3I_i\Om/J$. Angular-momentum conservation, including the torque required to maintain donor synchronization, is
\begin{align}
 \dot J&=-jm+N-\gamma_{\rm GW}J-\frac{d}{dt}(I_{\rm d}\Om),\nonumber\\
 \frac{d}{dt}(I_{\rm a}\omega_{\rm a})&=jm-N.
 \label{eq:budget}
\end{align}
The stream deposits specific angular momentum $j$ in the accretor, while the explicit donor spin derivative accounts for orbital exchange required by synchronization. Equation~(\ref{eq:budget}) conserves $J+I_{\rm d}\Om+I_{\rm a}\omega_{\rm a}$ subject to the gravitational radiation loss. We first derive the synchronized donor limit and subsequently extend the system to evolve the donor tidal response dynamically.

Let $x=\ln(R_{\rm d}/R_L)$ be the overfill coordinate and define $H=\dot a/(2a)$, $U=1-\epsilon_{\rm d}$, and
\begin{align}
 h_m&=\frac{1-q}{\Md}-\frac{j}{J}+\frac{\Om I'_{\rm d}}{J},\nonumber\\
 H&=\frac{h_m m+N/J-\gamma_{\rm GW}}{U}.
 \label{eq:H}
\end{align}
Combining Eqs.~(\ref{eq:roche}) and (\ref{eq:H}), we derive the conservative evolution equations
\begin{align}
 \dot x&=\frac{2(1-q)-\K}{\Md}m-2H,\nonumber\\
 \dot\delta&=\frac{j-\omega_{\rm a} I'_{\rm a}}{I_{\rm a}}m
                 -\frac{N}{I_{\rm a}}+3\Om H.
 \label{eq:evolution}
\end{align}
Setting the overfill and spin lag derivatives to zero, we derive the stationary contact solution for the same spin lag and stellar configuration,
\begin{align}
 m_0&=\frac{2\gamma_{\rm GW}\Md}{\Dt},\nonumber\\
 \Dt&=(1-\epsilon_{\rm a}-\epsilon_{\rm d})\K
 +2(\epsilon_{\rm a}+\epsilon_{\rm d})(1-q)\nonumber\\
 &\quad+\frac{2\Md}{J}(\Om I'_{\rm d}-\omega_{\rm a}I'_{\rm a}),\nonumber\\
 N_{\rm req}&=m_0\left[j-\omega_{\rm a}I'_{\rm a}\right.\nonumber\\
 &\quad\left.+\frac{3I_{\rm a}\Om}{2\Md}\bigl(2(1-q)-\K\bigr)\right].
 \label{eq:joint}
\end{align}
The limit $I_{\rm d}=I'_{\rm d}=0$ recovers the accretor spin contact model, and $I'_{\rm a}=0$ gives the constant inertia specialization. The donor spin contribution follows directly from the conservation equations.

\section{Local Dynamical Stability}
\begin{figure}[t]
\centering
\makebox[\columnwidth][c]{%
  \vbox{%
    \hbox{\includegraphics[width=.96\columnwidth]{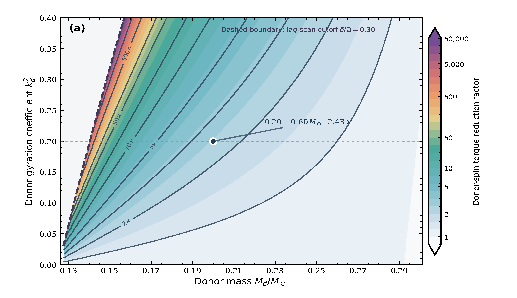}}%
    \nointerlineskip
    \hbox{\includegraphics[width=.96\columnwidth]{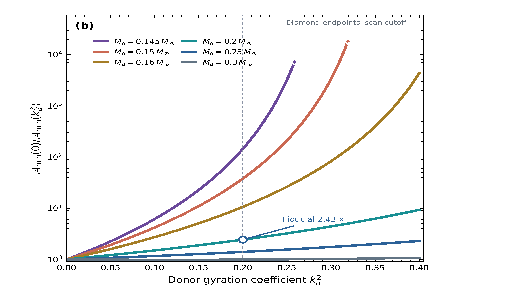}}%
  }%
}
\caption{Mass-dependent shift of the local torque threshold due to synchronized donor spin, calculated using 13-packet direct-impact streams for $M_{\rm a}=0.60\,\Mo$. Panel (a) maps the factor $\Aa_{\min}(0)/\Aa_{\min}(k_{\rm d}^2)$ over donor mass and squared gyration coefficient; contours mark factors of $2$, $5$, and higher where present. Panel (b) gives four mass cuts normalized to the zero-donor-spin limit. The expanded low-mass range is limited to configurations with a positive marginal torque solution; the effect is strongest at low donor masses and becomes weak near $0.30\,\Mo$. These calculated thresholds apply to the specified fifth power comparison torque and establish targets for stellar tide calculations.}
\label{fig:donorspin}
\end{figure}

We next determine the response of a stationary binary to perturbations. Its restoring dynamics are governed by the local torque gradient relative to the rate at which mass transfer amplifies overfill. For a local nozzle law $m(x)$, write $m_x=(dm/dx)_0>0$, $k_N=(dN/d\delta)_0$, and $\lambda_{\rm a}=m_0I'_{\rm a}/I_{\rm a}$. Here $k_N$ is a dimensional torque stiffness, distinct from the squared gyration coefficients $k_i^2$; $p\equiv\K m_x/\Md$ denotes the overfill-rate scale. At fixed background masses and orbital geometry, the Jacobian of Eq.~(\ref{eq:evolution}) is
\begin{align}
 \mathbf L&=\begin{pmatrix} A_xm_x&B_xk_N\\E_xm_x&-Ck_N-\lambda_{\rm a}\end{pmatrix},\nonumber\\
 A_x&=\frac{2(1-q)-\K}{\Md}-\frac{2h_m}{U},\quad
 B_x=-\frac{2}{JU},\nonumber\\
 E_x&=\frac{j-\omega_{\rm a}I'_{\rm a}}{I_{\rm a}}
      +\frac{3\Om h_m}{U},\quad
 C=\frac{1}{I_{\rm a}}-\frac{3\Om}{JU}.
 \label{eq:matrix}
\end{align}
The characteristic equation is $s^2-(\operatorname{tr}\mathbf L)s+\det\mathbf L=0$. Direct expansion gives the identity
\begin{align}
 \operatorname{tr}\mathbf L&=A_xm_x-Ck_N-\lambda_{\rm a},\nonumber\\
 \det\mathbf L&=\frac{\Dt m_x}{I_{\rm a}\Md U}\,k_N
                         -A_xm_x\lambda_{\rm a},\nonumber\\
 C&=\frac{1-\epsilon_{\rm a}-\epsilon_{\rm d}}{I_{\rm a}U}.
 \label{eq:detidentity}
\end{align}
The necessary and sufficient local stability inequalities are $\operatorname{tr}\mathbf L<0$ and $\det\mathbf L>0$. Throughout our sampled cold white dwarf domain, $\Dt,U,C,A_x>0$ and $\lambda_{\rm a}<0$, so a positive restoring torque slope ensures a positive determinant. We therefore derive the operative stability frontier from the trace,
\begin{equation}
 k_N>k_{\rm crit}=\frac{A_xm_x-\lambda_{\rm a}}{C}.
 \label{eq:kcrit}
\end{equation}
At the marginal trace crossing, the positive determinant yields $s_\pm=\pm i\sqrt{\det\mathbf L}$, establishing an oscillatory local transition. Equation~(\ref{eq:detidentity}) also supplies the full trace and determinant test for more general stellar models. The zero spin and instantaneous synchronization limits reproduce the structural and accretor torque boundaries of Marsh et al.~\cite{Marsh2004}. We calculate the donor inertia dependence across $0.13$ to $0.30\,\Mo$ in Fig.~\ref{fig:donorspin}, finding a pronounced reduction in the required torque at low donor mass and a progressively weaker effect toward the upper end. The mass cuts and two dimensional atlas follow from the same derived stability criterion.

\section{Nozzle Physics and Torque Capacity}
The restoring condition acquires a direct physical scale through the sensitivity of mass transfer to Roche overfill. We calculate this sensitivity using a degenerate polytropic throat of index $n_{\rm p}=3/2$, the sonic-flow closure scales as $m\propto(\Delta\Phi)^{n_{\rm p}+3/2}=(\Delta\Phi)^3$, where $\Delta\Phi$ is the potential overfill. Mapping the potential margin locally to $x$ gives $m_x\simeq3m_0/x_0$ at equilibrium logarithmic overfill $x_0$. The polytropic nozzle provides a defined reference to atmosphere dependent prescriptions~\cite{Ritter1988} and hydrodynamic transonic calculations~\cite{Cehula2023}. We define $m_x=\mu_x m_{x,0}$ to isolate the sensitivity to the local derivative, where $\mu_x$ is dimensionless and $m_{x,0}$ is the reference polytropic value. This construction varies the restoring response at fixed stationary mass flux $m_0$.

We assess a computed torque $N(\delta)$ through its value $N_{\rm req}$ and derivative $k_{\rm crit}$ at the same contact state. For a transparent analytical comparison, we define the family
\begin{align}
 N(\delta)&=T_0\Aa u^n\chi^{-h},\quad
 T_0=\frac{G\Md^2}{a}\left(\frac{R_{\rm a}}{a}\right)^5,\nonumber\\
 u&=\frac{2\delta}{\omega_{\rm dyn}},\quad
 \chi=\frac{\Om+\delta}{\delta},\quad
 \omega_{\rm dyn}=\left(\frac{G\Ma}{R_{\rm a}^3}\right)^{1/2}.
 \label{eq:templates}
\end{align}
Here $\Aa>0$ is the dimensionless torque normalization, $n>0$ and $h\ge0$ are assumed exponents, and $R_{\rm a}$ is the accretor radius. The $n=5$ reference follows the approximate fifth-power frequency envelope found in dynamical white-dwarf tide calculations~\cite{FullerLai2014}, while the detailed tidal spectrum develops resonant structure. We use $(n,h)=(5,0),(6,0),(5,5)$ as analytical comparisons of torque shape. For $t=\delta/\Om>0$, differentiation at fixed orbital configuration gives the exact logarithmic slope
\begin{equation}
 \nu_{\rm eff}=\frac{d\ln N}{d\ln\delta}=n+\frac{h}{1+t},\qquad
 \frac{N_{\rm req}}{\Om t}\nu_{\rm eff}>k_{\rm crit}.
 \label{eq:nu}
\end{equation}
We define $\nu_{\rm crit}=k_{\rm crit}\delta/N_{\rm req}$ and locate the marginal lag $t_*$ from $\nu_{\rm eff}=\nu_{\rm crit}$ at the same contact state. This yields the minimum required normalization
\begin{equation}
 \Aa_{\min}=\frac{N_{\rm req}(t_*)/T_0}
 {[2\Om t_*/\omega_{\rm dyn}]^n[t_*/(1+t_*)]^h}.
 \label{eq:amin}
\end{equation}

\subsection{Quadrupolar Mode Torques and Contact Branches}
\begin{figure}[t]
\centering
\makebox[\columnwidth][c]{%
  \vbox{%
    \hbox{\includegraphics[width=.96\columnwidth]{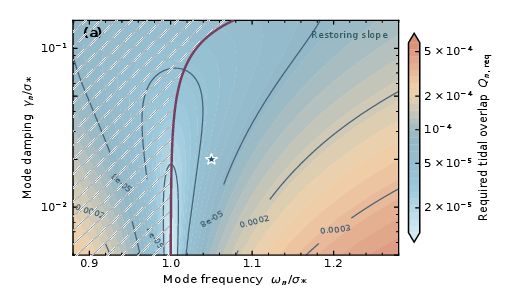}}%
    \nointerlineskip
    \hbox{\includegraphics[width=.96\columnwidth]{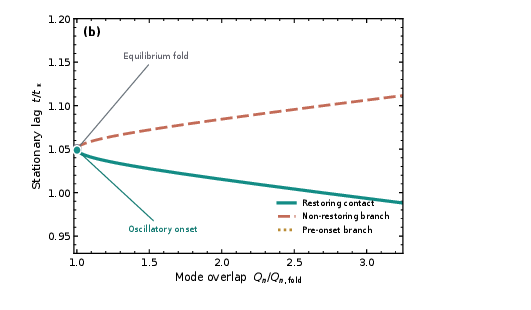}}%
  }%
}
\caption{Quadrupolar mode-torque stability frontier and stationary contact branches. (a) Required dimensionless tidal overlap $Q_{n,{\rm req}}$ for exact torque balance at the reference contact state, as a function of mode frequency and damping. The hatched domain fails the local restoring-gradient criterion; the solid mauve curve is $\nu_n=\nu_{\rm crit}$. The star marks the specified mode $\omega_n/\sigma_*=1.05$, $\gamma_n/\sigma_*=0.02$. (b) Paired torque-balanced stationary contacts for that same mode as its overlap varies. Solid teal denotes stable contact, dashed coral the unstable descending-torque branch, and dotted ochre the narrow pre-onset rising branch. The equilibrium fold and trace-controlled oscillatory threshold occur at different modal overlaps. The response is evaluated with the dimensional mode torque and the full local Jacobian.}
\label{fig:mode}
\end{figure}
A dimensional stellar tide connects the stability frontier to measurable mode properties. We evaluate a linearly damped quadrupolar mode using the established secular torque response~\cite{Burkart2013}
\begin{align}
 \frac{N_n(\sigma)}{T_0}&=16W_{22}^2 Q_n^2\frac{\omega_n^2\sigma\gamma_n}
 {(\omega_n^2-\sigma^2)^2+4\sigma^2\gamma_n^2},\nonumber\\
 \sigma&=2\delta,\qquad W_{22}^2=3\pi/10.
 \label{eq:physicalmode}
\end{align}
We define $Q_n$ as the dimensionless mode overlap, with $\omega_n$ and $\gamma_n$ denoting its eigenfrequency and damping rate. Differentiation gives the local logarithmic restoring slope
\begin{equation}
 \nu_n=1-\frac{4\sigma^2(\sigma^2-\omega_n^2+2\gamma_n^2)}{(\omega_n^2-\sigma^2)^2+4\sigma^2\gamma_n^2}.
 \label{eq:modeslope}
\end{equation}
At the reference contact frequency $\sigma_*=2\Omega t_*=1.19286\times10^{-3}\,{\rm rad\,s^{-1}}$, the amplitude condition $N_n(\sigma_*)=N_{\rm req}$ fixes an explicit required overlap,
\begin{equation}
 Q_{n,{\rm req}}^2=\frac{N_{\rm req}}{16W_{22}^2T_0}
 \frac{(\omega_n^2-\sigma_*^2)^2+4\sigma_*^2\gamma_n^2}
 {\omega_n^2\sigma_*\gamma_n}.
 \label{eq:qreq}
\end{equation}
The simultaneously required restoring condition is $\nu_n(\sigma_*)>\nu_{\rm crit}(t_*)=5$. Figure~\ref{fig:mode}(a) maps both constraints, retaining the absolute wave-torque normalization $N_{\rm req}/T_0=8.085\times10^{-8}$. For the explicitly specified mode $\omega_n/\sigma_*=1.05$ and $\gamma_n/\sigma_*=0.020$, the reference equilibrium requires $Q_n=5.425\times10^{-5}$ and satisfies the restoring criterion. The required overlap and damping therefore provide a direct quantitative benchmark for stellar mode calculations.

At fixed $\omega_n,\gamma_n,Q_n$, the physical stationary contacts solve $N_n(2\Omega t)=N_{\rm req}(t)$, with $N_{\rm req}(t)$ recomputed from the same conservative contact equations. Our stationary branch calculation in Fig.~\ref{fig:mode}(b) resolves two solutions. The equilibrium fold occurs at $Q_{n,{\rm fold}}=2.02098\times10^{-5}$ and $t=0.0215859$; the rising-torque branch crosses the oscillatory stability threshold at $Q_{n,{\rm osc}}=2.02352\times10^{-5}$ and $t=0.0215653$. At $Q_n=5.425\times10^{-5}$, two stationary states occur at $t=0.020562$ and $0.022610$, with $\max\operatorname{Re}(s)/p=-1.047$ and $+3.703$, respectively. The descending torque branch has a negative determinant, while the rising restoring branch becomes stable at its trace crossing. We thus resolve the torque amplitude threshold, the stationary fold, and the oscillatory frontier as distinct properties of the same specified tide. At the trace crossing, $\operatorname{Im}(s)/p=0.694666$ for $p=3.57440\times10^{-9}\,\mathrm{s^{-1}}$, corresponding to a secular oscillation period of $80.2$ yr. This calculated eigenperiod characterizes the local contact dynamics and provides a timescale for future coupled tidal and gravitational wave calculations. Stellar eigenfunctions determine the mode overlaps and damping rates appropriate to each accretor.

Our torque criterion makes the stellar physics requirement explicit. For the three minute reference contact, an extrapolated CO6 standing wave fit supplies a peak torque of $0.056N_{\rm req}$~\cite{Burkart2013}, yielding an eighteenfold amplitude requirement relative to that single mode. The derived frontier supplies a quantitative target for additional stellar modes and nonlinear traveling wave channels~\cite{Yu2020}, whose combined amplitude and local derivative determine the restoring response.

\begin{figure}[tb]
\centering
\makebox[\columnwidth][c]{%
  \vbox{%
    \hbox{\includegraphics[width=.96\columnwidth]{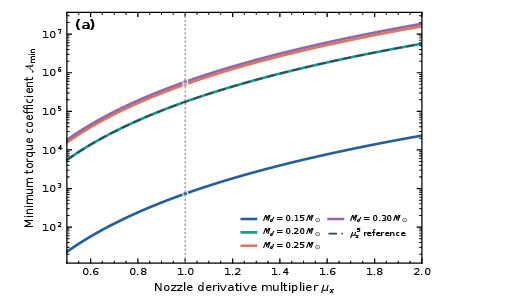}}%
    \nointerlineskip
    \hbox{\includegraphics[width=.96\columnwidth]{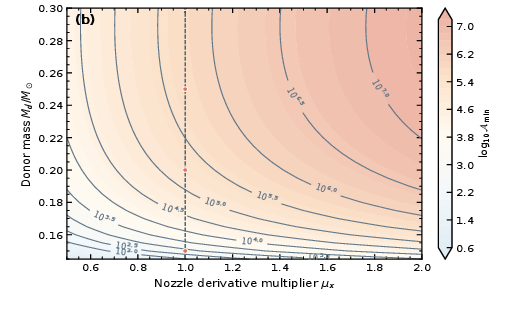}}%
  }%
}
\caption{Nozzle-gradient amplification of the marginal fifth-power torque requirement for $k_{\rm d}^2=0.20$ and $M_{\rm a}=0.60\,\Mo$. Panel (a) resolves four donor-mass sequences as the derivative multiplier $\mu_x$ varies, with the $\mu_x^5$ reference shown at $M_{\rm d}=0.20\,\Mo$. Panel (b) maps the corresponding absolute $\log_{10}\Aa_{\min}$ and contours across donor mass. Each mass uses an independently calculated 13 packet impact stream, with $\mu_x$ varied at fixed background equilibrium to isolate the derivative sensitivity.}
\label{fig:nozzle}
\end{figure}

We evaluate the reference fifth power torque at the same equilibrium with evolving stellar inertia and within the accretor breakup limit. Figure~\ref{fig:nozzle} then quantifies the nozzle derivative dependence across four donor masses.

\section{White-Dwarf Benchmarks and Uncertainty Propagation}
We now calculate the stability frontier for cold white dwarf binaries with consistently specified structure, overflow, and stream dynamics. The numerical model combines the Nauenberg mass and radius relation~\cite{Nauenberg1972}, Eggleton Roche geometry, a polytropic $L_1$ nozzle, and restricted three body direct impact trajectories. We derive the stream angular momentum from a 13 node velocity quadrature and adopt $M_{\rm a}=0.60\,\Mo$, $k_{\rm a}^2=0.18$, and $0\leq k_{\rm d}^2\leq0.20$, where zero is a comparison limit and $0.20$ is representative of a low-mass degenerate donor~\cite{SynchronousWD2009}. The baseline holds $k_{\rm a}^2$ constant, with its logarithmic mass derivative varied for structural sensitivity. For the $0.20+0.60\,\Mo$ benchmark, $\K=0.5972$, $\Dt=0.7080$, $m_0=1.053\times10^{17}\,\mathrm{kg\,s^{-1}}$, and $m_x=2.380\times10^{21}\,\mathrm{kg\,s^{-1}}$; the marginal values are $t_*=0.02056$, $N_{\rm req}=2.608\times10^{30}\,\mathrm{N\,m}$, $k_{\rm crit}=2.186\times10^{34}\,\mathrm{N\,m\,s}$, and $\Aa_{\min}=1.765\times10^5$.

As a structural check, the formalism recovers $q_{\rm crit}=0.6339795$ for the constant donor slope $\zeta_{\rm d}=-1/3$ and $q_{\rm crit}=0.5793833$ with the Nauenberg relation and a $0.60\,\Mo$ accretor. These calculated limits establish agreement with the classical Roche geometry.

\begin{table}[b]
\caption{Marginal fifth-power torque normalization and spin lag from the 13-packet direct-impact model. The two columns for donor spin compare $k_{\rm d}^2=0$ with $k_{\rm d}^2=0.20$.}
\label{tab:results}
\begin{ruledtabular}
\begin{tabular}{c r r r r}
$M_{\rm d}/\Mo$ & $\Aa_{\min}(0)$ & $\Aa_{\min}(.20)$ & $t_*(0)$ & $t_*(.20)$\\
\hline
0.15 & $2.798\!\times\!10^4$ & $7.36\!\times\!10^2$ & 0.04287 & 0.08778\\
0.20 & $4.292\!\times\!10^5$ & $1.765\!\times\!10^5$ & 0.01752 & 0.02056\\
0.25 & $6.978\!\times\!10^5$ & $4.987\!\times\!10^5$ & 0.01240 & 0.01288\\
0.30 & $5.745\!\times\!10^5$ & $5.803\!\times\!10^5$ & 0.01100 & 0.01040
\end{tabular}
\end{ruledtabular}
\end{table}

\begin{figure}[t]
\centering
\makebox[\columnwidth][c]{%
  \vbox{%
    \hbox{\includegraphics[width=.96\columnwidth]{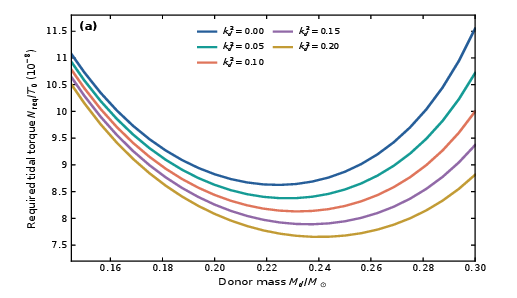}}%
    \nointerlineskip
    \hbox{\includegraphics[width=.96\columnwidth]{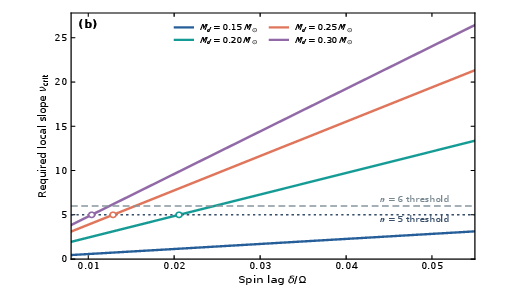}}%
  }%
}
\caption{Joint amplitude and differential-slope demands on a stabilizing tidal torque. Panel (a) gives the dimensionless return-torque requirement $N_{\rm req}/T_0$ for five donor-spin models. Panel (b) gives the required logarithmic stiffness $\nu_{\rm crit}=k_{\rm crit}\delta/N_{\rm req}$ for four donor masses at $k_{\rm d}^2=0.20$. The dotted $n=5$ and dashed $n=6$ lines are analytical comparison slopes; open circles mark the marginal fifth-power intersections within the plotted lag range. A computed stellar torque must satisfy magnitude and slope at the same spin lag.}
\label{fig:capacity}
\end{figure}

Our $0.20+0.60\,\Mo$ calculation gives a donor spin reduction of the required return torque from $2.846\times10^{30}$ to $2.608\times10^{30}\,\mathrm{N\,m}$ and a factor of $2.43$ change in the reference fifth power normalization. The factor approaches $38$ at $0.15\,\Mo$ as the marginal lag moves from $0.04287$ to $0.08778$, and crosses toward a small reversal near $0.30\,\Mo$ (Table~\ref{tab:results}). At fixed equilibrium, doubling the nozzle derivative increases the required normalization by $32$, consistent with the derived $\mu_x^5$ scaling. Changing the logarithmic inertia derivatives by $\pm0.30$ moves the reference threshold across $1.277\times10^5$ to $2.383\times10^5$, around the baseline $1.765\times10^5$. Figure~\ref{fig:capacity} displays the resulting amplitude and local stiffness requirements.

We verify the derived stability equations in their classical limits and by independent numerical differentiation. With negligible stellar inertia and complete orbital return of stream angular momentum, the conservative Roche condition $\K>0$ follows. The zero return torque limit yields $\dot x=2\gamma_{\rm GW}-[\K/\Md-2j/J]m$ and the direct impact boundary $\K>2\Md j/J$, recovering the limiting results of Marsh et al.~\cite{Marsh2004}. Our determinant identity agrees with direct evaluation to a relative precision of $4.24\times10^{-15}$, while 150 finite difference Jacobians agree to $6.41\times10^{-11}$. Increasing the stream quadrature from nine to thirteen packets at $0.20\,\Mo$ shifts $\Aa_{\min}$ from $175898.166$ to $176500.278$, a $0.341\%$ difference relative to the thirteen packet value. Direct eigenvalue calculations independently reproduce the marginal stability thresholds.

\begin{figure}[t]
\centering
\makebox[\columnwidth][c]{%
  \vbox{%
    \hbox{\includegraphics[width=.96\columnwidth]{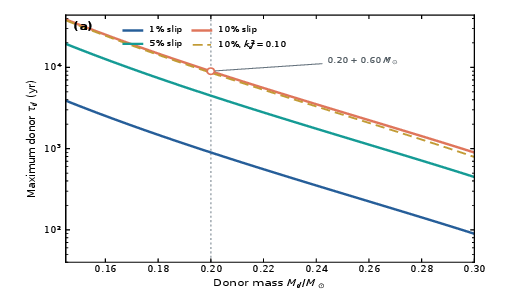}}%
    \nointerlineskip
    \hbox{\includegraphics[width=.96\columnwidth]{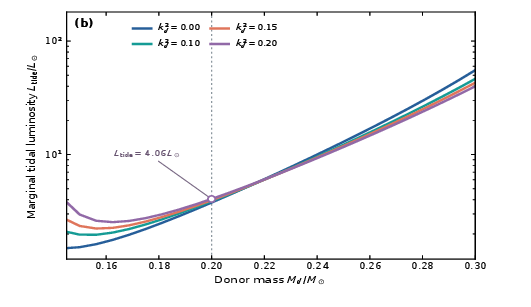}}%
  }%
}
\caption{Synchronized-donor and thermal feasibility requirements for the white-dwarf sequence. Panel (a) gives the upper limit on donor synchronization time for allowed spin slips of $1\%$, $5\%$, and $10\%$ with $k_{\rm d}^2=0.20$, alongside a $10\%$ reference for $k_{\rm d}^2=0.10$; the $0.20+0.60\,\Mo$ benchmark is marked explicitly. Panel (b) shows the marginal tidal-heating requirement for four donor inertias and highlights the fiducial heating benchmark $L_{\rm tide}\simeq4.06L_\odot$. These are conditional energetic and steady-slip demands, distinct from the finite-time stability boundary in Fig.~\ref{fig:finite_donor}.}
\label{fig:feasibility}
\end{figure}

\begin{figure}[t]
\centering
\makebox[\columnwidth][c]{%
  \vbox{%
    \hbox{\includegraphics[width=.96\columnwidth]{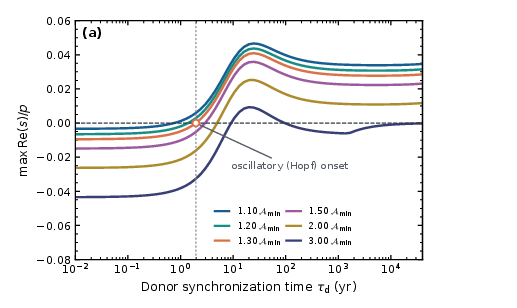}}%
    \nointerlineskip
    \hbox{\includegraphics[width=.96\columnwidth]{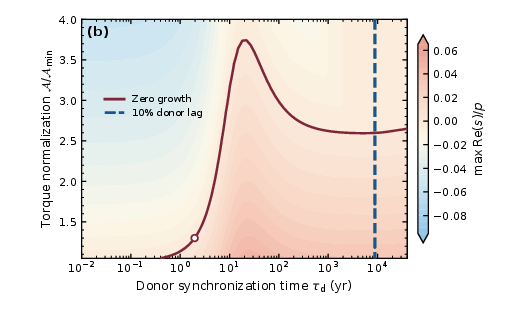}}%
  }%
}
\caption{Finite donor-tide stability in the $0.20+0.60\,\Mo$ reference model. Panel (a) compares six torque normalizations with the largest real part of the three-variable eigenvalue spectrum, normalized by $p=\K m_x/\Md$, against donor synchronization time. Panel (b) maps stable and growing regimes using pale contours. The solid curve marks zero growth and the dashed curve marks $10\%$ donor asynchronism; the oscillatory (Hopf) onset at $\tau_{\rm d}=1.958$ yr and $1.30\Aa_{\min}$ is indicated in both panels. The separate boundaries reveal that a nearly synchronous donor can still participate in an oscillatory instability. The benchmark uses the 13-packet stream and a specified linear donor tidal torque.}
\label{fig:finite_donor}
\end{figure}

\section{Synchronization and Thermal Feasibility}
The torque frontier also determines the synchronization and thermal demands placed on the stars. To maintain $\omega_{\rm d}=\Om$, the donor tide must balance the evolution of its spin angular momentum. The stationary contact equations fix $H_0=[2(1-q)-\K]m_0/(2\Md)$, from which we derive the required donor torque
\begin{equation}
 T_{{\rm d},{\rm req}}=\left|\frac{d}{dt}(I_{\rm d}\Om)\right|
 =\left|3I_{\rm d}\Om H_0+\Om I'_{\rm d}m_0\right|.
 \label{eq:tdonor}
\end{equation}
For a linear donor synchronizing torque $I_{\rm d}(\Om-\omega_{\rm d})/\tau_{\rm d}$ and an allowed fractional asynchronism $\varepsilon_{\rm slip}=|\omega_{\rm d}-\Om|/\Om$, the locked-donor assumption requires
\begin{equation}
 \tau_{\rm d}\le\varepsilon_{\rm slip}\frac{I_{\rm d}\Om}{T_{{\rm d},{\rm req}}}.
 \label{eq:sync}
\end{equation}
For the $0.20+0.60\,\Mo$ reference with $k_{\rm d}^2=0.20$, we obtain $T_{{\rm d},{\rm req}}=1.712\times10^{29}\,\mathrm{N\,m}$ and a $10\%$ steady slip bound of $\tau_{\rm d}\lesssim8.97\times10^3$ yr under the adopted linear donor tide (Fig.~\ref{fig:feasibility}). We calculate the accretor tidal heating as $L_{\rm tide}=N_{\rm req}\delta\simeq4.06L_\odot$ at the marginal fifth power contact. At this state, $m_0=1.053\times10^{17}\,\mathrm{kg\,s^{-1}}=1.67\times10^{-6}M_\odot\,\mathrm{yr}^{-1}$ and $GM_{\rm a}m_0/R_{\rm a}\simeq2.53\times10^3L_\odot$. The calculated orbital return torque is $N_{\rm req}=2.608\times10^{30}\,\mathrm{N\,m}$, corresponding to $N_{\rm req}/T_0=8.085\times10^{-8}$. These results place the restoration, synchronization, and heating requirements on a common physical scale.

\section{Finite Donor Synchronization}
We extend the synchronized donor model by promoting the donor spin lag $\Delta_{\rm d}=\omega_{\rm d}-\Om$ to a dynamical variable. This extension retains the conservative stream and accretor torque prescriptions while resolving the donor's finite tidal relaxation time. For $T_{\rm d}=-I_{\rm d}\Delta_{\rm d}/\tau_{\rm d}$ with $\tau_{\rm d}>0$, orbital angular momentum conservation and the donor spin equation yield
\begin{align}
 H&=h_bm+N/J+\frac{I_{\rm d}\Delta_{\rm d}}{J\tau_{\rm d}}-\gamma_{\rm GW},\nonumber\\
 \dot\Delta_{\rm d}&=-\frac{\Delta_{\rm d}}{\tau_{\rm d}}
 +\frac{\omega_{\rm d}I'_{\rm d}}{I_{\rm d}}m+3\Om H,\nonumber\\
 h_b&=\frac{1-q}{\Md}-\frac{j}{J}.
 \label{eq:donorfinite}
\end{align}
We retain the overfill and accretor lag relations in Eq.~(\ref{eq:evolution}), with $H$ specified by Eq.~(\ref{eq:donorfinite}). For the enlarged state $(x,\delta,\Delta_{\rm d})$, we derive the characteristic polynomial of the three dimensional Jacobian $\mathbf L_3$
\begin{align}
 P(s)&=s^3+a_1s^2+a_2s+a_3,\nonumber\\
 a_1&=-\operatorname{tr}\mathbf L_3,\quad a_3=-\det\mathbf L_3,\nonumber\\
 a_2&=\tfrac12\!\left[(\operatorname{tr}\mathbf L_3)^2
                         -\operatorname{tr}(\mathbf L_3^2)\right].
 \label{eq:cubic}
\end{align}
The Routh and Hurwitz inequalities $a_1,a_2,a_3>0$ and $a_1a_2>a_3$ provide the necessary and sufficient local stability test. In the strong coupling limit, one eigenvalue approaches $-U/\tau_{\rm d}$ while the other two recover Eq.~(\ref{eq:matrix}). We thereby identify a third eigenmode and an oscillatory boundary governed by donor relaxation, in addition to the accretor and orbital stability frontier. The physical torque and overflow equations determine the subsequent nonlinear response.

For the $0.20+0.60\,\Mo$ reference and the 13 packet stream, our cubic eigenvalue calculation places the oscillatory onset at $\tau_{\rm d}=1.958$ yr for a fifth power torque of $1.30\,\Aa_{\min}$, when the equilibrium donor lag is $2.18\times10^{-5}\Om$. At $3\,\Aa_{\min}$, the calculation yields a finite unstable interval from $9.15$ to $95.7$ yr. We verify the stability boundary with direct eigenvalues across the full $45\times80=3{,}600$ configuration grid. At long relaxation times the accretor and orbital eigenvalues approach the donor decoupled threshold $\mathcal A_{\min}(k_{\rm d}^2=0)=2.43\mathcal A_{\min}(k_{\rm d}^2=0.20)$ for the adopted fifth power torque. The oscillatory threshold for $1.30\Aa_{\min}$ occurs more than three orders of magnitude below the steady $10\%$ spin slip bound, revealing a distinct dynamical constraint even for nearly synchronized donors. This derived timescale identifies the stellar coupling required to restore contact after oscillatory perturbations~\cite{Gokhale2007,Toubiana2024}.

\begin{figure}[t]
\centering
\makebox[\columnwidth][c]{%
  \vbox{%
    \hbox{\includegraphics[width=.96\columnwidth]{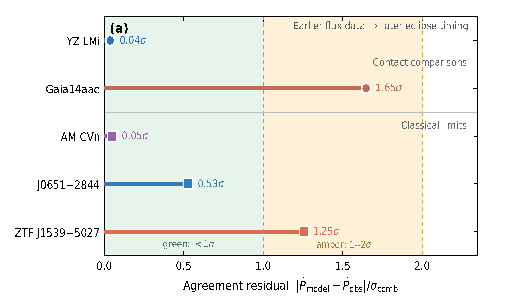}}%
    \nointerlineskip
    \hbox{\includegraphics[width=.96\columnwidth]{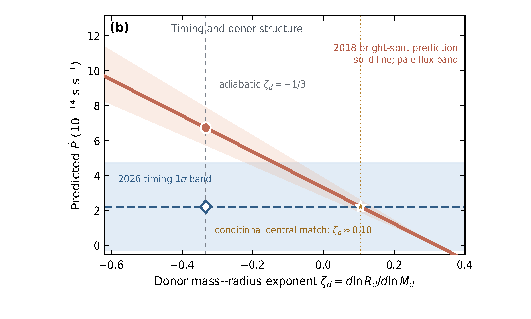}}%
  }%
}
\caption{Contact-kinematic and classical-limit consistency comparisons. (a) Five-binary agreement ladder using the combined prediction-and-measurement uncertainty $\sigma_{\rm comb}$. YZ LMi and Gaia14aae are pre-specified contact predictions from the 2018 bright-spot transfer rates and eclipse-derived masses; AM~CVn, J0651+2844, and ZTF~J1539+5027 are conservative or detached limiting-case benchmarks. All five comparisons have normalized residuals below $1.7\sigma$, supporting the contact kinematic benchmarks that complement the calculated direct impact stability frontier. (b) Gaia14aae donor-structure sensitivity. The solid curve and coral band show the predicted $\dot P$ and its 1$\sigma$ flux-driven range as a function of the donor mass--radius exponent $\zeta_{\rm d}$; the blue band shows the 2026 eclipse-timing measurement. The vertical dashed line marks the adiabatic reference $\zeta_{\rm d}=-1/3$, while the dotted line and star indicate the $\zeta_{\rm d}$ value that would match the observed central timing value. Gaia14aae illustrates conditional sensitivity of the contact relation to donor structure.}
\label{fig:observations}
\end{figure}

\section{Orbital Evolution and Observational Benchmarks}
The stationary contact solution yields an orbital timing relation that tests the kinematic component of the model. For conservative transfer with $m=-\dot M_{\rm d}>0$ and $\zeta_{\rm d}=d\ln R_{\rm d}/d\ln M_{\rm d}$, we derive the period evolution from Eq.~(\ref{eq:evolution}) as
\begin{equation}
\begin{aligned}
 \frac{\dot P}{P}&=\frac{3m}{2\Md}\,\mathcal B,\\
 \mathcal B&=2(1-q)-\K=-\zeta_{\rm d}+(1+q)f_q.
\end{aligned}
 \label{eq:chirp}
\end{equation}
The factor $3/2$ follows directly from $P\propto a^{3/2}$. This derived relation isolates the conservative Roche contact kinematics and provides an observational check complementary to the independently calculated dynamical stability frontier.

Using the published 2018 mass transfer estimates and independently measured eclipse masses of YZ LMi and Gaia14aae~\cite{Ramsay2018,Copperwheat2011,Green2018}, we calculate $\dot P=(3.77,6.72)\times10^{-14}\,\mathrm{s\,s^{-1}}$ for $\zeta_{\rm d}=-1/3$. The subsequent eclipse timing values, $(3.6\pm2.5,2.2\pm2.5)\times10^{-14}\,\mathrm{s\,s^{-1}}$~\cite{Green2026}, give combined uncertainty residuals of $0.04$ and $1.65$ standard deviations. We also recover the semi degenerate AM CVn contact limit and the detached quadrupole limits of J0651+2844 and ZTF J1539+5027 (Table~\ref{tab:orbital_benchmarks})~\cite{Smak2023,Hermes2012,Burdge2019}. Figure~\ref{fig:observations} displays the five comparisons and their combined uncertainties. For ZTF J1539 the difference corresponds to $1.25$ combined standard deviations once the propagated stellar mass uncertainty is included. The comparison establishes agreement of the derived contact kinematics with the available timing measurements and classical gravitational radiation limits.

\begin{table}[!ht]
\scriptsize
\caption{Orbital-period-derivative checks in units of $10^{-12}\,\mathrm{s\,s^{-1}}$. The first two rows compare contact-kinematic predictions with low-significance eclipse timing; the remaining rows check conservative or detached classical limits. Combined errors include estimated input and observational uncertainties.}
\label{tab:orbital_benchmarks}
\begin{ruledtabular}
\begin{tabular}{l r r r r}
System & Calculated & Observed & $\sigma_{\rm comb}$ & $|\Delta|/\sigma$ \\
\hline
YZ LMi & $+0.0377$ & $+0.036\pm0.025$ & $0.047$ & $0.04$ \\
Gaia14aae & $+0.0672$ & $+0.022\pm0.025$ & $0.027$ & $1.65$ \\
\hline
AM CVn & $+0.878$ & $+0.85\pm0.48$ & $0.569$ & $0.05$ \\
J0651+2844 & $-8.19$ & $-9.8\pm2.8$ & $3.05$ & $0.53$ \\
ZTF J1539 & $-21.83$ & $-23.73\pm0.05$ & $1.51$ & $1.25$ \\
\end{tabular}
\end{ruledtabular}
\end{table}

For fixed Gaia14aae masses and transfer rate, the central measured $\dot P$ corresponds to $\zeta_{\rm d}=0.105$, while its timing $1\sigma$ interval spans $-0.138\lesssim\zeta_{\rm d}\lesssim0.347$ before propagation of transfer rate uncertainty. Our calculation in Fig.~\ref{fig:observations}(b) quantifies how improved timing can probe donor structure. The disc accreting contact systems provide kinematic benchmarks, while our direct impact calculations deliver the complementary torque stability frontier. The numerical data and source code underlying the figures are available from the author upon reasonable request.

\section{Conclusions}
The dynamical survival of interacting binary stars is governed by the restoring response of the coupled orbit and stellar spins, beyond the balance that establishes Roche contact. We derive a conservation based stability model incorporating Roche overflow, the angular momentum carried by the accretion stream, evolving stellar inertias, and the spins of both stars. The stationary equations determine the return torque required for mass transfer, and an exact Jacobian identity yields the independent restoring gradient. In our computed direct impact white dwarf domain this reduces to a trace controlled oscillatory frontier. Extending the model to finite donor synchronization reveals a second dynamical boundary associated with the donor spin response.

Our dimensional quadrupolar tide calculation resolves the stationary equilibrium fold at $Q_{n,{\rm fold}}=2.02098\times10^{-5}$ and the oscillatory stability onset at $Q_{n,{\rm osc}}=2.02352\times10^{-5}$, thereby separating the existence of contact from its local stability. For the $0.20+0.60\,M_\odot$ reference, we obtain a factor of $2.43$ shift in the fifth power torque threshold from donor spin, a strong dependence on the nozzle derivative, and an oscillatory onset at $\tau_{\rm d}=1.958$ yr for $1.30\mathcal A_{\min}$. The resulting hierarchy of equilibrium, restoring, and synchronization thresholds is the central mathematical advance of the model.

The observational comparison provides a complementary test of the theory. The orbital period calculations agree with the published conservative contact and detached gravitational wave benchmarks within combined uncertainties, while the derived torque thresholds supply independent targets for stellar tide calculations. In the reference state, the extrapolated CO6 single mode torque reaches $0.056N_{\rm req}$, establishing an eighteenfold amplitude requirement for additional tidal channels. For a binary with $P=216.6\,\mathrm{s}$ and $f_{\rm GW}=9.23\,\mathrm{mHz}$, component masses and orbital timing constrain the contact geometry and donor response, while mode calculations can test the amplitude and restoring stiffness demanded by local stability. By deriving these complementary constraints in a single conservation framework, we establish a quantitative route from stellar tidal physics to the survival and secular evolution of interacting compact binaries in the millihertz gravitational wave population.


\begin{thebibliography}{0}%
\makeatletter
\providecommand \@ifxundefined [1]{%
 \@ifx{#1\undefined}
}%
\providecommand \@ifnum [1]{%
 \ifnum #1\expandafter \@firstoftwo
 \else \expandafter \@secondoftwo
 \fi
}%
\providecommand \@ifx [1]{%
 \ifx #1\expandafter \@firstoftwo
 \else \expandafter \@secondoftwo
 \fi
}%
\providecommand \natexlab [1]{#1}%
\providecommand \enquote  [1]{``#1''}%
\providecommand \bibnamefont  [1]{#1}%
\providecommand \bibfnamefont [1]{#1}%
\providecommand \citenamefont [1]{#1}%
\providecommand \href@noop [0]{\@secondoftwo}%
\providecommand \href [0]{\begingroup \@sanitize@url \@href}%
\providecommand \@href[1]{\@@startlink{#1}\@@href}%
\providecommand \@@href[1]{\endgroup#1\@@endlink}%
\providecommand \@sanitize@url [0]{\catcode `\\12\catcode `\$12\catcode
  `\&12\catcode `\#12\catcode `\^12\catcode `\_12\catcode `\%12\relax}%
\providecommand \@@startlink[1]{}%
\providecommand \@@endlink[0]{}%
\providecommand \url  [0]{\begingroup\@sanitize@url \@url }%
\providecommand \@url [1]{\endgroup\@href {#1}{\urlprefix }}%
\providecommand \urlprefix  [0]{URL }%
\providecommand \Eprint [0]{\href }%
\providecommand \doibase [0]{http://dx.doi.org/}%
\providecommand \selectlanguage [0]{\@gobble}%
\providecommand \bibinfo  [0]{\@secondoftwo}%
\providecommand \bibfield  [0]{\@secondoftwo}%
\providecommand \translation [1]{[#1]}%
\providecommand \BibitemOpen [0]{}%
\providecommand \bibitemStop [0]{}%
\providecommand \bibitemNoStop [0]{.\EOS\space}%
\providecommand \EOS [0]{\spacefactor3000\relax}%
\providecommand \BibitemShut  [1]{\csname bibitem#1\endcsname}%
\let\auto@bib@innerbib\@empty
\end{thebibliography}%


\begin{thebibliography}{20}
\footnotesize
\bibitem{Eggleton1983} P. P. Eggleton, Astrophys. J. \textbf{268}, 368 (1983).
\bibitem{Nauenberg1972} M. Nauenberg, Astrophys. J. \textbf{175}, 417 (1972).
\bibitem{Ritter1988} H. Ritter, Astron. Astrophys. \textbf{202}, 93 (1988).
\bibitem{Marsh2004} T. R. Marsh, G. Nelemans, and D. Steeghs, Mon. Not. R. Astron. Soc. \textbf{350}, 113 (2004).
\bibitem{Burkart2013} J. Burkart, E. Quataert, P. Arras, and N. N. Weinberg, Mon. Not. R. Astron. Soc. \textbf{433}, 332 (2013).
\bibitem{FullerLai2014} J. Fuller and D. Lai, Mon. Not. R. Astron. Soc. \textbf{444}, 3488 (2014).
\bibitem{Cehula2023} J. Cehula and O. Pejcha, Mon. Not. R. Astron. Soc. \textbf{524}, 471 (2023).
\bibitem{Yu2020} H. Yu, N. N. Weinberg, and J. Fuller, Mon. Not. R. Astron. Soc. \textbf{496}, 5482 (2020).
\bibitem{Gokhale2007} V. Gokhale, X. M. Peng, and J. Frank, Astrophys. J. \textbf{655}, 1010 (2007).
\bibitem{Toubiana2024} A. Toubiana, N. Karnesis, A. Lamberts, and M. C. Miller, Astron. Astrophys. \textbf{692}, A165 (2024).
\bibitem{SynchronousWD2009} W. Even and J. E. Tohline, Astrophys. J. Suppl. Ser. \textbf{184}, 248 (2009), doi:10.1088/0067-0049/184/2/248.
\bibitem{Ramsay2018} G. Ramsay \textit{et al.}, Astron. Astrophys. \textbf{620}, A141 (2018).
\bibitem{Copperwheat2011} C. M. Copperwheat \textit{et al.}, Mon. Not. R. Astron. Soc. \textbf{410}, 1113 (2011).
\bibitem{Green2026} M. J. Green \textit{et al.}, Mon. Not. R. Astron. Soc. \textbf{548}, stag673 (2026); correction \textbf{550}, stag1011 (2026).
\bibitem{Green2018} M. J. Green \textit{et al.}, Mon. Not. R. Astron. Soc. \textbf{476}, 1663 (2018).
\bibitem{Smak2023} J. Smak, Acta Astron. \textbf{73}, 227 (2023), doi:10.32023/0001-5237/73.3.2.
\bibitem{Hermes2012} J. J. Hermes \textit{et al.}, Astrophys. J. Lett. \textbf{757}, L21 (2012).
\bibitem{Burdge2019} K. B. Burdge \textit{et al.}, Nature \textbf{571}, 528 (2019).
\end{thebibliography}
\end{document}